# Simulation computation in grammar-compressed graphs

Stefan Böttcher, Rita Hartel and Sven Peeters

*Paderborn University*
*Fürstenallee 11*
*Paderborn, 33102, Germany*
*{stb@, rst@, speeters@campus.}uni-paderborn.de*

*Abstract*: Like [1], we present an algorithm to compute the simulation of a query pattern in a graph of labeled nodes and unlabeled edges. However, our algorithm works on a compressed graph grammar, instead of on the original graph. The speed-up of our algorithm compared to the algorithm in [1] grows with the size of the graph and with the compression strength.

## 1. Introduction

### 1.1. Motivation

A fundamental operation on large graphs is the search of occurrences of a given pattern in the graph. Like [1], we consider graphs and patterns of labeled nodes and unlabeled edges and compute a relation called *simulation of the pattern in the graph*, which determines a superset of the occurrences of the pattern in the graph. An algorithm in [1] computes a simulation of a pattern within a graph in time $O(n*m)$ for n vertices and m edges of the graph. This algorithm starts with a huge set of candidate nodes in the graph for each node of the pattern, and it sharpens this set of candidate nodes with each edge of the query pattern being processed. Especially, when the pattern occurs frequently in the graph, this algorithm can have a long runtime as processing an edge to a node of the pattern requires processing numerous edges to numerous candidate nodes in the graph.

In contrast, our algorithm works on grammar-compressed graphs, i.e. computes the simulation of a pattern on a compressed graph grammar. This reduces the data volume processed in the sharpening steps from huge sets of candidate nodes to significantly smaller sets of so-called grammar path suffixes, where each grammar path suffix represents a compressed set of candidate nodes. The larger the graph and the stronger the compression, the greater is the speed-up of our algorithm in comparison to the algorithm in [1]. Our approach requires only one initial compression of a huge graph into a compressed graph grammar. Thereafter, we can compute the simulation of arbitrary query patterns on this compressed graph grammar.

### 1.2. The original pattern search algorithm

[1] defines a candidate occurrence of a pattern graph *Q* in a given graph *OG* as follows.

Let $OG=(V_G, E_G, \lambda_G)$, $Q=(V_Q, E_Q, \lambda_Q)$, $V_G$ and $V_Q$ be node sets, $E_G$ and $E_Q$ be edge sets, and $\lambda_G$ and $\lambda_Q$ be functions mapping nodes to labels. *Q* can occur in *OG* iff there exists a binary simulation relation $SIM \subseteq V_Q \times V_G$, such that

1. for each $(u,v) \in SIM$, $\lambda_Q(u) = \lambda_G(v)$ holds
2. for each $u' \in V_Q$ there exists $v' \in V_G$, such that
   (a) $(u',v') \in SIM$, and
   (b) for each $(u',u) \in E_Q$, there exists $(v',v) \in E_G$, such that $(u,v) \in SIM$

[1] provides an algorithm to compute *SIM* on a graph *OG*. We explain the algorithm in [1] using the original graph *OG* shown in Figure 1(a) and the pattern graph *Q* that we are searching for shown in Figure 1(c). For each node $u \in V_Q$, a set $sim[u] \subseteq V_G$ of candidate nodes is computed and is repeatedly *sharpened* by removing nodes that violate condition 2(b). Initially, for each $u \in V_Q$, the set of candidate nodes *sim[u]* is preset with all nodes of *OG* carrying the same label as *u*, e.g., for the node $u \in V_Q$ with label *c*, *sim[u]*={1,3,6,8}, and for the node $u' \in V_Q$ with label *d*, *sim[u']*={2,4,7,9}. The algorithm in [1] achieves its goal, i.e. to *sharpen* the node set *sim[u']*, by repeating three further steps. First, it computes the set of all predecessors *pre(sim[u])*={2,5,7} of nodes of set *sim[u]*. Second, it computes the set *remove*={1,3,4,6,8,9} of nodes that violate condition 2(b), i.e. nodes of the old set of predecessors *old_pre(sim[u])* (which initially is set to all nodes of the graph) that are no predecessors of *sim(u)*. Third, these nodes are removed from *sim[u']*, i.e., the new set *sim[u']* contains the nodes of the old set *sim[u']* that are not contained in the *remove* set, i.e., nodes 2 and 7. Thereafter, other node sets *sim[u]* with $u \in V_Q$ are sharpened, until a fixed point is reached for all sets *sim[u]*. In this example, finally, *sim[u]*={6} and *sim[u']*={7}, i.e. *SIM*={(u,6),(u',7)}.

## 2. Pattern search on graph grammars

We only once compress the original graph *OG* to a graph grammar *GG* of *OG*. Then, our algorithm can compute a simulation relation for arbitrary pattern queries *Q* on *GG*. In comparison to the algorithm of [1], for this purpose, our algorithm regards rather small sets of grammar path suffixes in *GG* instead of using huge sets of nodes in *OG*.

We first introduce our approach to grammar-based graph compression with examples and formal definitions, before we describe our simulation computation algorithm.

### 2.1. Our approach to grammar-based graph compression

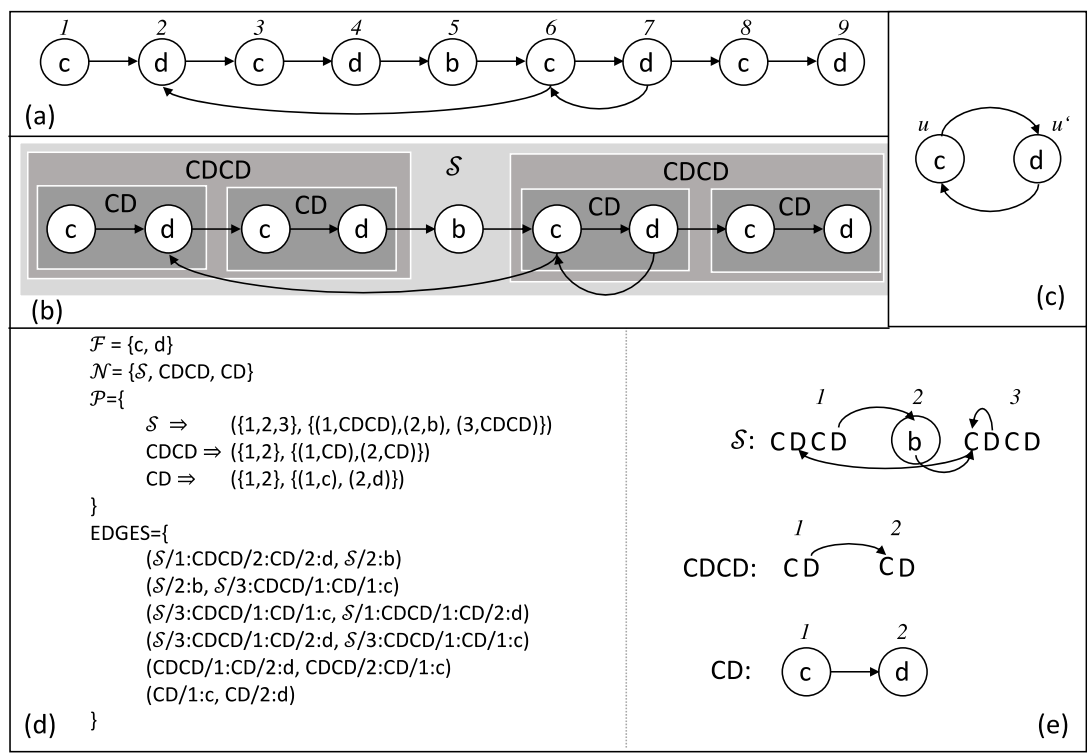

*Figure 1: (a) Original Graph OG, (b) stepwise generation of GG from OG, (c) pattern graph Q, (d) compressed graph grammar GG for OG, and (e) GG's visualization.*

Figure 1(a) shows the original graph *OG* used in our example, Figure 1(d) shows the compressed graph grammar *GG* of *OG,* and Figure 1(e) shows a visualization of *GG*.

*GG* can be generated stepwise from *OG=($V_G$, $E_G$, $\lambda_G$)* as shown in Figure 1(b). Each (inner) rectangle surrounds a subgraph that occurs multiple times. For each such subgraph, a rule is added to *GG*. And each occurrence of the subgraph is replaced by a nonterminal for that rule. In each compression step, we select and replace a subgraph occurring most frequently.

In a first compression step (visualized by the innermost rectangles shown in Figure 1(b)), each occurrence of a subgraph containing an edge *($v_1,v_2$)* with *$\lambda(v_1)$=c* and *$\lambda(v_2)$=d* within *OG* is replaced with a nonterminal node *CD*, and the rule for *CD* is added to the grammar in Figure 1(d). Thereby, the four edges *($v_1,v_2$)* with *$\lambda(v_1)$=c* and *$\lambda(v_2)$=d* in *OG* are compressed to the single edge *(CD/1:c,CD/2:d)* listed in the bottom of Figure 1(d). This edge is visualized in Figure 1(e) as an edge from node *1*, labeled *c*, to node *2*, labeled *d* in the grammar rule for *CD*. In a second compression step, each occurrence of a subgraph containing an edge *($v_1,v_2$)* with *$\lambda(v_1)$=CD* and *$\lambda(v_2)$=CD* in the graph resulting from the first step is replaced with a nonterminal *CDCD*, and the rule for *CDCD* is added to the grammar in Figure 1(d). Thereby, two edges *($v_1,v_2$)* with *$\lambda(v_1)$=CD* and *$\lambda(v_2)$=CD* within in the graph resulting from the first compression step are compressed to the single edge *(CDCD/1:CD/2:d,CDCD/2:CD/1:c)*. This edge is visualized in Figure 1(e) as an edge from node *1*, labeled *CD*, to node *2*, also labeled *CD* in the grammar rule for *CDCD*.

When there are no more subgraphs occurring multiple times, the remaining graph forms the right-hand side of the so-called *start rule* with nonterminal *S,* and the remaining edges are added to the set *EDGES* of *GG*.

While the set of all 4 *d*-labeled nodes of *OG* is compressed into a single *d*-labeled node in the *CD* rule of *GG*, we can represent each single node of *OG* by a so-called grammar path. For example, a grammar path *gp0=S/3:CDCD/1:CD/2:d* denotes a sequence of calls, from the node with ID 3 in the *S* rule via the node with ID 1 in the *CDCD* rule to the node with ID 2 in the *CD* rule labeled *d*. We say that *gp0 represents* the node set *{7}⊂$V_G$*, or *rep(gp0)={7}* for short. Furthermore, a *grammar path suffix*, e.g. *gps1=CDCD/1:CD/2:d*, *represents* the set of all nodes that are represented by a grammar path of which *gps1* is a suffix, here, the node set *rep(gps1)={2,7}⊂$V_G$*. Similarly, the grammar path suffix *gps2=CDCD/2:CD/2:d* represents the node set *rep(gps2)={4,9}⊂$V_G$*, and the grammar path suffix *gps3=d* represents all the nodes in *OG* with label *d*, i.e., the node set *{2,4,7,9}⊂$V_G$*. Finally, the pair of grammar path suffixes *(CD/1:c,CD/2:d) represents* all the edges from a node with label *c* to a node with label *d* in *OG*, i.e., the edge set *{(1,2),(3,4),(6,7),(8,9)}⊂$E_G$*.

Similar to [2], our approach to compression selects and replaces in each compression step a subgraph containing only one edge and occurring most frequently. However, in comparison to [2], we can compress cyclic graphs instead of just trees, and we use a completely different technique to represent edges which is tailored to our pattern search.

Formally, an alphabet $\Sigma$ = $\mathcal{N}$ ∪ $\mathcal{F}$ is the union of two disjoint finite sets: a set $\mathcal{N}$ of non-terminal symbols and a set $\mathcal{F}$ of terminal symbols.

Let *OG = (V, E, $\lambda$)* be a directed graph, where *V* is a strictly ordered node set and *ord(v)*∈ $\mathbb{N}$ denotes the ordinal number of *v* for each *v*∈*V*. Furthermore *E*⊆*V*x*V* is an edge set and each node *v*∈*V* has a label *$\lambda(v)$*∈ $\mathcal{F}$.

A linear context-free graph grammar is a 5-tuple $GG = (\mathcal{F},\mathcal{N},\mathcal{P},\mathcal{S},EDGES)$, where $\mathcal{F}$ is the set of terminal symbols, $\mathcal{N}$ is the set of nonterminal symbols, $\mathcal{F}\cap\mathcal{N} = \emptyset$, $\mathcal{P}$ denotes the set of rules, and $\mathcal{S}\in\mathcal{N}$ is the start nonterminal symbol which does not occur on the right-hand side of any rule. For each $N \in \mathcal{N}$, there is exactly one rule $(N \Rightarrow rhs(N)) \in \mathcal{P}$, with $rhs(N)=(V, \lambda)$ where $V$ is an ordered set of nodes and $\lambda: V \rightarrow \Sigma$ is a label function. Finally, $EDGES \subseteq GPS \times GPS$ is a global set of pairs $(N/i:gps1,N/j:gps2)$ of so-called grammar path suffixes, both starting with the same nonterminal $N$. Each *grammar path suffix* $gps \in GPS$ is a string $N_1/i_1:...:N_n/i_n:N_{n+1}$ with $N_{n+1} \in \mathcal{F}$ and for each $k \in \{1,...,n\}$ holds: $N_k \in \mathcal{N}$ and there is a node $v \in rhs(N_k)$, such that $ord(v)=i_k$ and $\lambda(v)=N_{k+1}$. Furthermore, $n \geq 0$, i.e., the shortest grammar path suffix consists of a label $F \in \mathcal{F}$ only.

Whenever a grammar path suffix *gps* starts with $N_1 = \mathcal{S}$, we call it a grammar path *gp*.

We call $GG=(\mathcal{F},\mathcal{N},\mathcal{P},\mathcal{S},EDGES)$ a *graph grammar of* $OG=(V_G, E_G, \lambda_G)$ if $GG$ can be decompressed to $OG'=(\mathcal{F},\mathcal{N},\{\mathcal{S} \Rightarrow (V_G,\lambda_G)\},\mathcal{S},E')$ with $E'=\{ (\mathcal{S}/i:\lambda_G(i), \mathcal{S}/j:\lambda_G(j)) \mid (i,j) \in E_G\}$ by repeatedly *inlining into* all nonterminal nodes of rule $\mathcal{S}$, as follows. (We consider non-recursive grammars only, i.e., grammars that can be decompressed into a unique finite graph $OG'$.)

Let $v \in V_S$ be a non-terminal node of $rhs(\mathcal{S})=(V_S,\lambda_S)$ with ordinal number $ord_S(v)$ and $N=\lambda_S(v) \in \mathcal{N}$. Furthermore, let the nodes $v_i \in V_N$ of $rhs(N)=(V_N, \lambda_N)$ have the ordinal numbers $ord_N(v_i)$. And, let $\oplus$ denote the string concatenation symbol. To *inline into* $v$ means that we add copies $c_i$ of the nodes $v_i \in V_N$ to the nodes of $V_S$ and remove $v$, i.e., $V_S=(V_S \cup copyOf(V_N))-\{v\}$. Added nodes $c_i \in copyOf(V_N)$ get the new ordinal numbers $ord_S(c_i)=ord_N(v_i)+max(\{ord_S(v)|v \in V_S\})$, but keep their labels, i.e., $\lambda_S(c_i)=\lambda_N(v_i)$. Furthermore, for each grammar path $gp$ of the form $\mathcal{S}/ord_S(v):\lambda_S(v)/i_2:\oplus gps$, we replace $(gp, gpx) \in EDGES$ with $(gp', gpx) \in EDGES$ and $(gpx, gp) \in EDGES$ with $(gpx, gp') \in EDGES$, where $gp'=\mathcal{S}/(maxOrdS+i_2):\oplus gps$. Finally, for each grammar path suffix $gpse= \lambda_S(v)/i_2:\oplus gps$, such that there is a grammar path suffix $gpx$ with $(gpse, gpx) \in EDGES$ or $(gpx,gpse) \in EDGES$, an additional edge $e=(gpse',pgx)$ (or $e=(gpx, gpse')$ respectively) with $gpse'=\mathcal{S}/(maxOrdS+i_2):\oplus gps$ is added to $EDGES$.

Note that there is a one-to-one mapping $rep_o: GP \rightarrow V_G: rep_o(gp)=v$ between the grammar paths $GP$ contained in a graph grammar $GG$ of an original graph $OG$ and the nodes $V_G$ of $OG$. We define this one-to-one mapping $rep_o$ from a grammar path $gp=\mathcal{S}/i_1:N_1/.../i_{n+1}:F$ to the node $v \in V_G$ with $\lambda(v)=F$ that $gp$ represents by a sequence of inlining steps. Within each step, we inline into the nonterminal $N_1$ of $gp$ until, finally, $gp=\mathcal{S}/n:F$. Then, $rep_o(gp)$ is that node $v \in OG$ with $ord_\mathcal{S}(v)=n$.

Let $gps$ be a *grammar path suffix* and $gp_1,...,gp_n$ be all grammars paths of which $gps$ is a suffix, then we say, $gps$ *represents* the set $rep(gps):=\{rep_o(gp_1),...,rep_o(gp_n)\}$ of nodes. And a set $GPS=\{gps_1, ..., gps_n\}$ of grammar path suffixes represents all the nodes which are represented by any $gps_i \in GPS$, i.e., $REP(GPS) := rep(gps_1) \cup ... \cup rep(gps_n)$.

### 2.2. Simulation computation on graph grammars

Our main contribution is that we reduce pattern simulation on a given huge graph $OG$ to pattern simulation on an often significantly smaller graph grammar $GG$. For this purpose, we represent huge sets of nodes $V_G$ of $OG$ by sets of grammar path suffixes, and we use the advantage that a single grammar path suffix summarizes multiple grammar paths in $GG$ and thereby represents many nodes in $V_G$. That is, in comparison to the algorithm in [1], we do not compute sets of candidate nodes, but sets of candidate grammar path suffixes, i.e., in our optimized simulation implementation, the sets

$sim_g[u]$, $sim_g[u']$, and $remove_g$, are each a set of grammar path suffixes. Furthermore, in each step of our algorithm, these sets of grammar path suffixes represent the corresponding node sets of the algorithm in [1], i.e., $REP(sim_g[u]) = sim[u]$, $REP(sim_g[u']) = sim[u']$, $REP(remove_g) = remove$, etc.

Before we outline our algorithm, we continue using the previous example.

Initially, the sets $sim_g[u]$ and $sim_g[u']$ are sets of grammar path suffixes containing only the label *c* and *d* respectively, and the set of old predecessors contains all terminal nodes of *GG* only. To sharpen $sim_g[u']$ by computing predecessors of $sim_g[u]$, our algorithm uses only one grammar path suffix *c* in $sim_g[u]$, instead of the four *c*-labeled nodes 1,3,6,8 used by the algorithm in [1]. While the algorithm in [1] keeps nodes 2 and 7 of *OG* in *sim[u']*, as they are *d*-labeled predecessor nodes of the four *c*-labeled nodes 1,3,6,8 in *sim[u]*, we search a set of grammar path suffixes that represents exactly these two *d*-labeled predecessor nodes. As *{d}* represents all four *d*-labeled nodes, we split *{d}* into the equivalent set *{ CDCD/**1**:CD/2:d , CDCD/**2**:CD/2:d }* of two grammar path suffixes. As only the first grammar path suffix, *CDCD/**1**:CD/2:d,* represents the node set *{2,7}*, only this grammar path suffix is added to $pre_g(sim_g[u])$.

Further predecessors of $sim_g[u]=\{c\}$ are the grammar path suffixes *S/2:b* and *gps2=S/3:CDCD/1:CD/2d*, both of which do not need to be considered further for different reasons. *S/2:b* represents *b*-labeled nodes $v \in V_G$, i.e. does not sharpen $sim_g[u']$ which represents *d*-labeled nodes $v \in V_G$. And *gps1* is a suffix of *gps2*. Therefore, each node represented by *gps2* is also represented by *gps1*, such that we do not have to add *gps2* to $pre_g(sim_g[u])$ in addition to *gps1*. We also say that *gps2* is *subsumed* by *gps1*.

Our algorithm continues on *GG* similarly as the algorithm in [1] works on *OG*, and finally, the new set $sim_g[u']$ is {*CDCD/**1**:CD/2:d*}. Note however that our algorithm operates on sets of grammar path suffixes that are significantly smaller than the node sets on which the algorithm in [1] operates. In the example, the *remove* set for the graph of Figure 2(a) contains 6 nodes, whereas the $remove_g$ set computed for the compressed graph grammar contains only the 2 grammar path suffixes *{c, CDCD/2:CD/2:d}*.

In general, our algorithm shown in Figure 2 computes a simulation relation $SIM_g$ between a graph-grammar $GG=(\mathcal{F},\mathcal{N},\mathcal{P},\mathcal{S},EDGES)$ of the original graph $OG=(V_G,E_G,\lambda_G)$ and the query pattern $Q=(V_Q, E_Q, \lambda_Q)$ solely based on *GG* and on *Q*. Our simulation relation $SIM_g$ contains for each $u \in V_Q$, the set $sim_g[u]$ of those grammar path suffixes of *GG* that represent at least one node $v \in V_G$ with $(u,v) \in SIM$.

To calculate $SIM_g$, we use the sets $sim_g[u]$, $oldsim_g[u]$, and $old\_pre\_of\_sim_g[u]$ of grammar path suffixes for each node $u \in V_Q$. Intuitively, each grammar path suffix in $sim_g[u]$ represents nodes $v \in V_G$ that are candidates for a simulation for *u*, i.e. for $(u,v) \in SIM$. The sets $oldsim_g[u]$ and $old\_pre\_of\_sim_g[u]$ are used for tracking the last states of $sim_g[u]$ and $pre_g(sim_g[u])$ and to allow an efficient *sharpening* of $sim_g[u']$.

Initially (lines 2-4), for each node $u \in V_Q$, the set $sim_g[u]$ of candidates contains only one grammar path suffix $\lambda_Q(u)$, which represents that each node $v \in V_G$ with $\lambda_G(v)=\lambda_Q(u)$ is a candidate node for *u*. Furthermore, $oldsim_g[u]$ and $old\_pre\_of\_sim_g[u]$ contain one grammar path suffix $\lambda_G(f)$ for each terminal symbol $f \in \mathcal{F}$ occurring in the grammar.

The loop consisting of lines 5-11 is called *sharpening*. In each iteration, we select one $u \in V_Q$ and its set $sim_g[u]$ of candidate grammar path suffixes for excluding candidate grammar path suffixes from other sets $sim_g[u']$ for predecessors $u' \in V_Q$ of *u* as follows.

```
GLOBAL variables: PatternGraph Q=(VQ,EQ, λQ), GraphGrammar GG=(𝓕,…,EDGES)

PROCEDURE simulate
1.  for u ∈ VQ do:
2.      oldsim_g[u] = 𝓕.to_grammar_path_suffixes
3.      old_pre_of_sim_g[u] = 𝓕.to_grammar_path_suffixes
4.      sim_g[u] = { λQ[u] | u in VQ }.to_grammar_path_suffixes
5.  while we find u ∈ VQ with sim_g[u] ≠ oldsim_g[u] do:
6.      oldsim_g[u] = sim_g[u]
7.      pre_of_sim_g_u = pre_g(sim_g[u])
8.      remove_g = δ(old_pre_of_sim_g[u],pre_of_sim_g_u)
9.      for each u' ∈ VQ with (u',u) ∈ EQ do:
10.         sim_g[u'] = δ(sim_g[u'],remove_g)
11.     old_pre_of_sim_g[u] = pre_of_sim_g_u
12. return (∀u∈VQ:sim_g[u]≠{}) ? {(u,gps)|∃u∈VQ ∃gps∈sim_g[u]} : {}

PROCEDURE pre( GPS )
13. pres = Union over (gps ∈ GPS) of preOFgps(gps)
14. return pres.sorted().remove_subsumed_paths

PROCEDURE preOFgps( gps )
15. return  { lgps | ∃ string prefix: (lgps,prefix⊕gps) ∈ EDGES }
16.       ∪ { prefix⊕lgps | ∃ string prefix, ∃ (lgps,rgps) ∈ EDGES:
                                                gps = prefix⊕rgps }

PROCEDURE δ(from,toRemove).
17. Ext = from
18. repeat
19.     if ∃ ext ∈ Ext, ∃ rem ∈ toRemove with rem is suffix of ext :
20.         Ext = Ext - {ext}
21.     else if ∃ ext ∈ Ext, ∃ rem ∈ toRemove with ext is suffix of rem :
22.         Ext = Ext ∪ OneStepExtensions(ext) - {ext}
23. until Ext does not change anymore in repeat-loop
24. return Ext

PROCEDURE OneStepExtensions(ext)
25. F = firstLabel(ext)
26. return { NT ⊕ "/" ⊕ NodeID ⊕ ":" ⊕ ext |
            ∃ NT ➔ (V_NT,E_NT, λ_NT) and λ_NT(NodeID)=F }
```

*Figure 2: The procedure* `simulate` *of our algorithm*

In line 5, we select a node $u \in V_Q$ for which $sim_g[u] \neq oldsim_g[u]$, i.e, a node for which grammar path suffixes have been removed from $sim_g[u]$ since the last selection of $u$. For the new set $sim_g[u]$ of such a node $u$, a new set $pre_g(sim_g[u])$ of predecessor grammar path suffixes is computed (line 7). In line 8, we compute the set $remove_g$ of grammar path suffixes that do no longer represent candidate nodes for the simulation, after $sim_g[u]$ has been sharpened. Each grammar path suffix $gps$ contained in the set $remove_g$ has to fulfill two conditions. First, $gps$ has to be subsumed by a grammar path suffix in the set $old\_pre\_of\_sim_g[u]$, i.e. the set $sim_g[u]$ of the previous iteration (line 11) must contain a suffix of $gps$. Second, $gps$ has to represent nodes in $V_G$, which violate condition 2(b), as these nodes are *not* a predecessor of a node of $V_G$ that simulates $u$, i.e., $gps$ must not represent a node in $V_G$ that is also represented by the new set $pre_g(sim_g[u])$. All the grammar path suffixes $gps$ contained in $remove_g$ are split off from and excluded from the grammar path suffixes of the sets $sim_g[u']$ for each predecessor node $u'$ of $u$ in $Q$ (lines 9-10).

Sharpening stops if no grammar path suffix of any set $sim_g[u]$ can be excluded anymore, i.e. $sim_g[u]=oldsim_g[u]$ for all $u \in V_Q$. Finally, in line 12, we check whether for each node $u \in V_Q$, $sim_g[u]$ is non-empty. If so, we return the set of all pairs (u,gps) where gps is a grammar path suffix of $sim_g[u]$. Note that we could instead return the larger set $SIM=\{(u,v)/\exists u \in V_Q \exists gps \in sim_g[u] \exists v \in rep(gps)\}$. Otherwise, we return the empty set.

Note that in our algorithm, no decompression of *GG* to *OG'* is needed, which keeps the search space smaller than in the algorithm of [1].

### 2.3. Predecessor sets of *sim* and of *oldsim*

The *predecessor set of a set of grammar path suffixes GPS* is defined as the union of the predecessor sets of all grammar path suffixes in *GPS*, i.e. $PRE_g(GPS)= \cup_{gps \in GPS} pre_g(GPS)$ (line 13). In order to return the smallest possible set (line 14), we remove all grammar path suffixes $sub \in pre_g(GPS)$ that are subsumed by other grammar path suffixes $sup \in pre_g(GPS)$, i.e., for which *sup* is a suffix of *sub*.

The predecessor set of a single grammar path suffix *gps* denoted by $pre_g(gps)$ is a set of non-overlapping grammar path suffixes, such that for each node $v' \in V_G$ that is a predecessor in *OG* of a node $v \in rep(gps)$, there exists exactly one grammar path suffix $gps' \in pre_g(gps')$ with $v' \in rep(gps')$.

Lines 15-16 show a simple way to compute the predecessor set of a single grammar path suffix. For $(lgps,rgps) \in EDGES$, (line 15) considers the case that *gps* is a suffix of *rgps*. Then, *gps* subsumes *rgps*, i.e., $rep(rgps) \subseteq rep(gps)$. In this case, $pre_g(gps)$ has to represent the start nodes *v'* of each edge $(v',v) \in E_G$, with $v \in rep(rgps)$. And *lgps* is exactly the grammar path suffix representing all the start nodes *v'* of these edges.

However, (line 16) considers the case that $gps=prefix \oplus rgps$, i.e., *rgps* is a suffix of *gps*. Then, *rgps* subsumes *gps*, i.e., $rep(gps) \subseteq rep(rgps)$. If $rep(gps) \neq rep(rgps)$, we do not want to represent all the predecessors *v'* of $v \in rep(rgps)$, but only predecessors *v'* of $v \in rep(gps)=rep(prefix \oplus rgps)$. That is why we only consider edges of $E_G$ represented by $(prefix \oplus lgps, prefix \oplus rgps)$. The start nodes *v'* of these edges are represented by $prefix \oplus lgps$, i.e. the grammar path suffix returned for $prefix \oplus rgps$ in line 16.

As a result of procedure `pre(GPS)` (lines 13-14), $pre_g(GPS)$ is a set of nonoverlapping grammar path suffixes representing precisely all predecessors of nodes $n \in REP(GPS)$.

### 2.4. Grammar path suffix difference set *δ*

In lines 8 and 10 of our algorithm, we compute a path difference set *δ(from,remove)* of two sets of grammar path suffixes *from* and *remove*. *δ(from,remove)* is a set of grammar path suffixes that represents of the set difference $REP(from)$ -$_{set}$ $REP(remove)$ of node sets. To compute the result *Ext* of *δ(from,remove)*, we start with the set *from* (line 17) and restrict this set step by step. Whenever there exists a pair of elements $ext \in Ext$ and $rem \in remove$, such that *ext* represents a subset of the nodes represented by *rem*, i.e. $rep(ext) \subseteq rep(rem)$, we have to remove *ext* from *Ext*. This is the case, if *rem* is a suffix of *ext* (line 19). Otherwise, if *rem* represents a proper subset of *ext*, i.e., $rep(rem) \subset rep(ext)$, *ext* is a suffix of *rem* (line 21). In this case, we have to find a set of grammar path suffixes that represents the set difference $rep(ext)$ -$_{set}$ $rep(rem)$ of node sets. For this purpose, we search for all "calls" of *ext*, i.e., all rules, the right-hand side of which contains a label that equals the first step of *ext*. For each such call, we create a grammar path suffix that extends *ext* by one step (lines 25-26). The union of all these extended grammar path suffixes then replaces *ext* in *Ext* (line 22). In later iterations through lines 19-22, these longer grammar path suffixes are again compared with the elements of the set *remove* and potentially deleted from *Ext*.

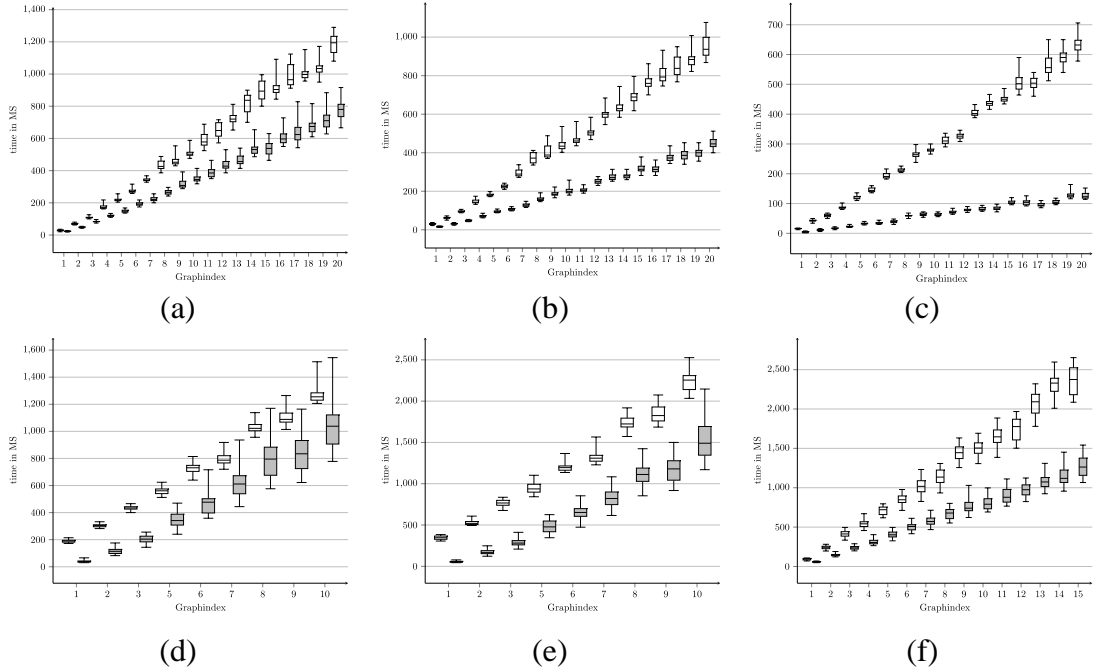

*Figure 3: Evaluation results for (a)-(c) random graphs, (d)-(e) social networks, and (f) RDF graphs*

### 2.5. Optimized implementation

*pre_g(GPS)* is calculated in a bottom-up fashion to avoid multiple inspections of the same rule. Furthermore, removing and splitting off of grammar path suffixes is delayed until really necessary to avoid unnecessary computations of a path difference set $\delta$.

## 3. Evaluation

We compared our approach for pattern search on compressed graphs with the algorithm in [1], running on a non-compressed graph. We performed two series of measurements.

Within the first series, we created random graphs with different compression ratios. Each graph consists of many variations of the same sub-graph, where for each variation, we delete randomly up to the half of the nodes of each sub-graph. In order to scale the compression ratio, we added random edges between these sub-graphs. Figure 3(a)-(c) show the results. The graphs scale from around 15,000 nodes (graphindex:1) up to around 300,000 nodes (graphindex:20). Figure 3(a) shows the results for graphs with approx. 2 edges per node on average, yielding compression ratios around 1:5, Figure 3(b) for graphs with approx. 1.6 edges per node (compression ratios around 1:7), and Figure 3(c) for graphs with approx. 1.25 edges per node (compression ratios around 1:20). The white box plots show the runtimes of the algorithm in [1], and the gray box plots show the runtimes of our algorithm. As we can see, our algorithm on compressed graphs clearly outperforms the algorithm [1], whereas the benefit is bigger the bigger the graphs get and the stronger the compression ratio becomes.

In a second series of measurements, we compared the two algorithms on the LDBC Social Network Benchmark and on a subgraph of dbpedia (nodes around "Angela Merkel") as an example for RDF graphs. The LDBC graphs vary from 100,000 nodes

(graphindex:1) to 515,000 nodes (graphindex:10) and the RDF graphs vary from 50,000 nodes (graphindex:1) to 730,000 nodes (graphindex:15). Figure 3(d) shows the results for LDBC for random pattern graphs with 6 nodes and 8 edges, and 3(e) for patterns with 10 nodes and 15 edges. Again, our algorithm outperforms the algorithm in [1], whereas the benefit increases for more complex patterns and bigger graphs. The same holds for the RDF graphs, as shown in Figure 3(f) for patterns with 8 nodes and 7 edges.

To summarize, our algorithm outperforms the algorithm [1] in all of our tests, and the benefit increases the bigger the graphs are, the stronger the compression is, and the more complex the pattern graphs are.

## 4. Related Work

There exist different approaches to compress graphs, but only for few of them there exist approaches to speed-up operations on the graphs by benefitting from compression.

The approaches presented in [3] and [4] compress web graphs by combining and efficiently encoding large sub-matrices of 1- or 0-bits within the adjacency matrix of webgraphs. For the k2-tree, [3] presents a compression algorithm for web graphs and presents an approach on how to navigate to the neighbours of a node within the k2-tree representation in linear time. Furthermore, [5] extends these ideas to algorithms for the set operations union, intersection, difference, and complement in linear time over the size of the k2-tree.

Grammar-based compression replaces repeatedly occurring sub-structures by nonterminal symbols and a rule, mapping the nonterminal to the replaced sub-structure, and it has been previously used for compressing strings [8], trees [2], and graphs [7].

An approach to the traversal problem for string grammars, i.e., extracting the next symbol without unnecessary decompression in constant time is presented in [9]. For tree grammars, this problem was studied in [10]. [11] presents a solution to this problem for straight-line hyperedge replacement grammars, i.e., the search for nodes connected to a given node through a (hyper)edge with a given label. In contrast to [7] and [11], we use grammar rules to compress multiple nodes instead of multiple edges/hyperedges, and we yield a completely different representation of edges which is optimized for pattern search as search operation.

Only few further approaches are known that use a compressed version of the graph to speed-up operations on the graph. [12] use a compressed graph to speed up link analysis, size estimations and several algorithms based on matrix-vector products for web graphs. [13] compresses graphs in such a lossy way that certain structural properties that are necessary to compute certain queries are kept. [14] uses compression to reduce the search space to speed-up the subgraph isomorphism problem.

Similar to all these approaches, we incorporate graph-compression in order to speed-up a search operation on the graph, which in our case is the pattern simulation. In contrast to [13] we use a lossless compression technique, such that the original graph can be restored from it, i.e., we do not need to preserve a non-compressed version of the graph. In contrast to [14] we are not restricted to combine nodes that have exactly the same outgoing edges. As a consequence, we can combine a larger set of nodes of the original graph into a single node of the compressed graph grammar, such that we can exclude larger set of nodes within a simulation computation step (i.e., we reduce the search space even stronger).

# 5. Conclusions

For computing the simulation of a search pattern query on graphs with labeled nodes and unlabeled edges, we have presented a simulation computation algorithm that operates on compressed graph grammars, instead of on the original graph. The speed-up of our algorithm gets greater, the larger the graph is and the stronger the compression is. The essence of our approach is to represent large node sets by significantly smaller sets of grammar path suffixes and to substitute the sharpening on these large node sets by a sharpening on the sets of grammar path suffixes. We assume that our approach, i.e. to perform algorithms on grammar path suffixes of graph grammars instead of on large node sets is applicable to far more search algorithms and even to algorithms that go beyond search.